\documentclass[%
aip,
graphicx,
amsmath,
amssymb,
reprint
]{revtex4-1}

\usepackage[dvipsnames]{xcolor}
\usepackage{hyperref}
\usepackage[utf8]{inputenc}
\usepackage[T1]{fontenc}
\usepackage{mathptmx}
\usepackage{etoolbox}

\usepackage{wrapfig}

\usepackage[normalem]{ulem}
\usepackage{graphicx}
\usepackage{soul}

\begin{document}

\title{Frequency modulated enhancement of microwave resonator sensing} 

\author{Pranaya Kishore Rath}
\email[]{rathpra1@msu.edu}
\affiliation{Department of Physics and Astronomy, 
Michigan State University, East Lansing MI 48824 USA}

\author{James D. Philips}
\email[]{jim.philips@zhinst.com}
\affiliation{Zurich Instruments, 400 5th Ave Waltham, MA 02451}

\author{Taekwan Yoon}
\affiliation{Zurich Instruments, 400 5th Ave Waltham, MA 02451}

\author{Kent R. Shirer}
\affiliation{Zurich Instruments, 400 5th Ave Waltham, MA 02451}

\author{Arash Fereidouni}
\affiliation{Zurich Instruments, 400 5th Ave Waltham, MA 02451}

\author{Johannes Pollanen}
\email[]{pollanen@msu.edu}
\affiliation{Department of Physics and Astronomy, 
Michigan State University, East Lansing MI 48824 USA}

\date{\today}

\begin{abstract}
We use the Pound-Drever-Hall (PDH) technique to characterize the frequency stability of a microwave-frequency surface acoustic wave (SAW) resonator-based sensor. 
The multi-mode acoustic resonator is integrated in a notch geometry with a transmission line, all fabricated on Y-cut lithium niobate. 
We measure the amplitude and phase of the resonator's transfer function and the PDH signal across the resonator's full spectral range. We use these measurements to emphasize the differences between the PDH measurement and a standard Phase-Locked Loop (PLL) technique. As compared to a PLL, we demonstrate that PDH is insensitive to phase error and exhibits a reduced Allan deviation of the center frequency measurement, in each case by up to an order of magnitude.
The method rejects spurious effects and background frequency drift, demonstrating the enhancements possible with PDH-based measurements, which can be realized in a wide range of microwave-frequency resonator-based sensors and devices.

\end{abstract}


\maketitle 

Microwave-frequency resonators have established themselves as effective sensors across diverse scientific and industry applications~\cite{Zmuidzinas2012,Nyfors2000,aslam2023advances}. 
By converting changes in a physical quantity into changes in frequency, they facilitate extremely precise and accurate measurements of parameters ranging from temperature~\cite{WANG2021}, pressure~\cite{IBRAHIM2011,XUE2022}, and strain~\cite{KELLY2022,Hosoi2025} to molecular composition~\cite{HOSSEINI2021,LEUSMANN2024}.  
The performance of these sensors primarily depends on their sensitivity to external stimuli and the precision in measuring key spectral parameters, such as center frequency and lineshape. 
A straightforward method to measure the center frequency $f_0$ involves driving the resonator with an oscillator frequency $f$, swept across the resonance using a network analyzer.  
The resonator’s amplitude and phase response can then be analyzed to extract $f_0$ and the linewidth.
For real-time sensing, a Phase-Locked Loop (PLL) is often employed, where the oscillator frequency is continuously adjusted to maintain a constant phase of the signal returning from the resonator~\cite{deBellescize1932}. 
PLLs are widely used in signal generators and analyzers, high-frequency stabilizers, and resonator-based sensors~\cite{gardner2005phaselock}. 
While effective in detecting the center frequency, PLLs are susceptible to phase errors from parasitic impedance and do not directly measure the spectral linewidth. In contrast, the Pound-Drever-Hall technique is well-established for stabilizing lasers to cavities and atomic transitions~\cite{Drever1983} and can overcome these limitations.  
Originally developed for microwave measurements~\cite{pound1946electronic}, PDH has been instrumental in several Nobel Prize-winning experiments. A notable example is its application within the Laser Interferometer Gravitational-Wave Observatory (LIGO)~\cite{black2001introduction}. 
Although the technique was first introduced for microwave systems, its use in modern microwave resonator applications has been limited~\cite{lindstrom2011pound}. 
In particular, it has not been explored for surface acoustic wave (SAW) resonator-based sensors nor for micro-electromechanical systems (MEMS) in general.

\begin{figure*}
    \centering
    \includegraphics[width=0.9\linewidth]{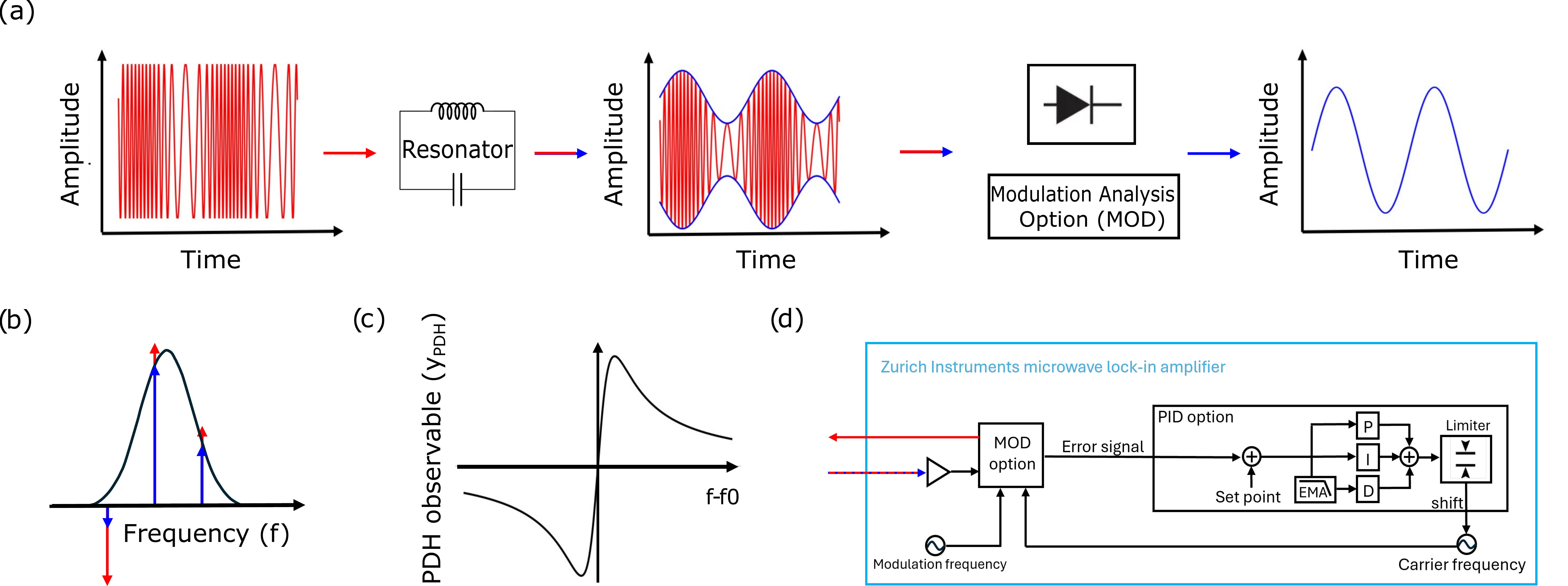}
      \caption{(a)~Time-domain representation of a frequency-modulated (FM) microwave input (red), detuned from the resonator center frequency, $f_0$, interacting with a resonator and resulting in amplitude modulation, (blue). Demodulation of the signal can be performed through a square-law detector, at $f_m$, or via the modulation analysis (MOD) option of a Zurich Instrument lock-in.(b)~Frequency-domain representation of the input FM signal (red) interacting with a resonator (black), resulting in an amplitude-modulated output (blue). (c)~PDH observable, $y_{PDH}$, versus frequency offset $f-f_0$, which can be used as an error signal in a feedback loop for resonance tracking. (d)~Functional diagram of the PDH feedback loop using a Zurich Instruments Microwave Lock-in: MOD extracts the error signal, PID minimizes it by adjusting the carrier frequency, and the updated FM drive is fed back to MOD.}
\end{figure*}

SAWs are mechanical oscillations confined to the surface of crystals, extending about one wavelength below the surface.  
On piezoelectric materials, SAWs can be generated using interdigitated transducers (IDTs), which convert electrical signals to mechanical waves. 
SAW devices, operated in the MHz-GHz range, are highly sensitive to changes in surface conditions, including pressure, temperature, and chemical species, enabling a wide range of sensing applications. Their fast response time, high sensitivity, non-contact operation, compact design, and low power consumption, further enhance their utility in both commercial and research applications~\cite{Morgan2007}. 
These characteristics have led to their widespread use in biosensing~\cite{huang2021surface,baumgartner2023recent,lange2008surface,fourati2023applications,pohanka2018overview,mujahid2019overview,hartz2020lateral,naranda2022practical,baumgartner2023recent,skladal2024piezoelectric}, chemical sensing~\cite{gomes2001application,kuchmenko2019perspective,li2023advances}, environmental monitoring~\cite{galipeau1991study,he2013high,wang2015surface,memon2022surface}, and industrial process control~\cite{aslam2023advances}. Additionally, their ability to be either capacitively or inductively coupled along with their intrinsically small spatial footprint also make SAW devices exciting for hybrid quantum technologies, where they have been integrated with superconducting qubits~\cite{manenti2017circuit, satzinger2018quantum,Moores2018,kitzman2023phononic, Undershute2025, andersson2019non}, semiconductor based quantum systems~\cite{Delsing2019} like quantum dots\cite{gell2008modulation,kataoka2007single,patel2024surface}, and solid state defect centers~\cite{Schuetz2015,whiteley2019spin}. Such piezoelectric coupling enables phonon-mediated quantum state transfer~\cite{mcneil2011demand, Hermelin2011,bertrand2016fast, Barnes2000, Byeon2021}, mechanically assisted quantum information processing~\cite{gustafsson2014propagating,dumur2021quantum,bienfait2019phonon}, and coherent microwave-to-optical transduction~\cite{Shumeiko2016}. Beyond information processing, SAW resonators may serve as quantum-enhanced sensors capable of detecting excitations within coupled qubit–phonon systems~\cite{kitzman2023quantum2,cleland2024studying}. As SAW-based quantum devices are anticipated to play a role in advancing next-generation quantum technologies and sensors continued improvements in resonator design, modeling, and real-time measurement techniques will further enhance their capabilities.

In this paper, we present PDH measurement of a SAW-based microwave resonator. We measure the amplitude and phase of the resonator transfer function and the PDH signal across the spectrum of the multi-mode SAW resonator, characterizing its resonances. We show that PDH is able to sensitively detect the effects of spurious resonator modes, which is a challenge for conventional techniques. PDH can also suppress the response to these spurious acoustic modes via a careful choice of the excitation signal. We show that PDH outperforms PLL by effectively suppressing parasitic influences in the measurement of the resonator frequency. PDH achieves a better Allan deviation, revealing that the resonator frequency is intrinsically more stable than a PLL indicates.

PDH sensing is carried out by driving the device under test with a carrier signal at frequency $f$, which is frequency modulated at $f_m$ (see Fig.~1~(a,b).
When the FM signal interacts with the resonator, it acquires amplitude modulation (AM) that depends on the offset of the carrier frequency $f$ from the resonator frequency, $f_0$.  
The AM signal is typically measured with a power detector. Demodulating this signal at $f_m$ yields the PDH observable, $y_{PDH}$, which enables an accurate measurement of $f-f_0$.

In the microwave domain, the FM signal generation has traditionally relied on analog modulators, while the resulting AM can be detected using analog power detectors. The relatively limited dynamic range of these components can challenge precise PDH measurements. With the advent of fully digital, FPGA-based lock-in amplifiers, implementing the PDH technique in the microwave regime has become considerably more straightforward and robust. The FM signal can be generated using multiple digital oscillators to produce the three-tone spectrum required for PDH, comprising of the carrier, lower sideband (LSB), and upper sideband (USB). The resulting AM can be extracted through two sequential demodulations (``double demodulation").
In this work, however, we detected each of the three tones (carrier, LSB, USB) independently, a method we refer to as direct detection. This method improves upon double demodulation, as it eliminates the bandwidth constraint on the first demodulation stage, allowing broader flexibility in the choice of modulation frequency $f_m$. 

We construct the amplitude of the PDH observable, $y_{PDH}$, from the in-phase ($X$) and quadrature ($Y$) components of the three demodulated tones (carrier, LSB, USB),
\begin{align}
y_{PDH,X} = X_0 (X_U+X_L )+Y_0 (Y_U+Y_L )\\
y_{PDH,Y} = Y_0 (X_L-X_U )-X_0 (Y_L-Y_U),
\end{align}
where $X_0$ ($Y_0$) is the in-phase (quadrature) component of the carrier tone, and $X_U$ ($Y_U$) and $X_L$ ($Y_L$) are the components of the USB and LSB, respectively~\cite{black2001introduction}. The experiments utilized a Zurich Instruments GHFLI, a fully digital, FPGA-based lock-in amplifier equipped with two signal inputs, eight oscillators, and eight demodulators, covering DC to $1.8$~GHz. The digital oscillators generated the FM spectrum, while the parallel demodulators enabled direct detection. This all-digital approach substantially enhances dynamic range and measurement stability compared to traditional analog modulation and detection schemes.

\begin{figure}[ht]
    \centering\includegraphics[width=0.8\linewidth]{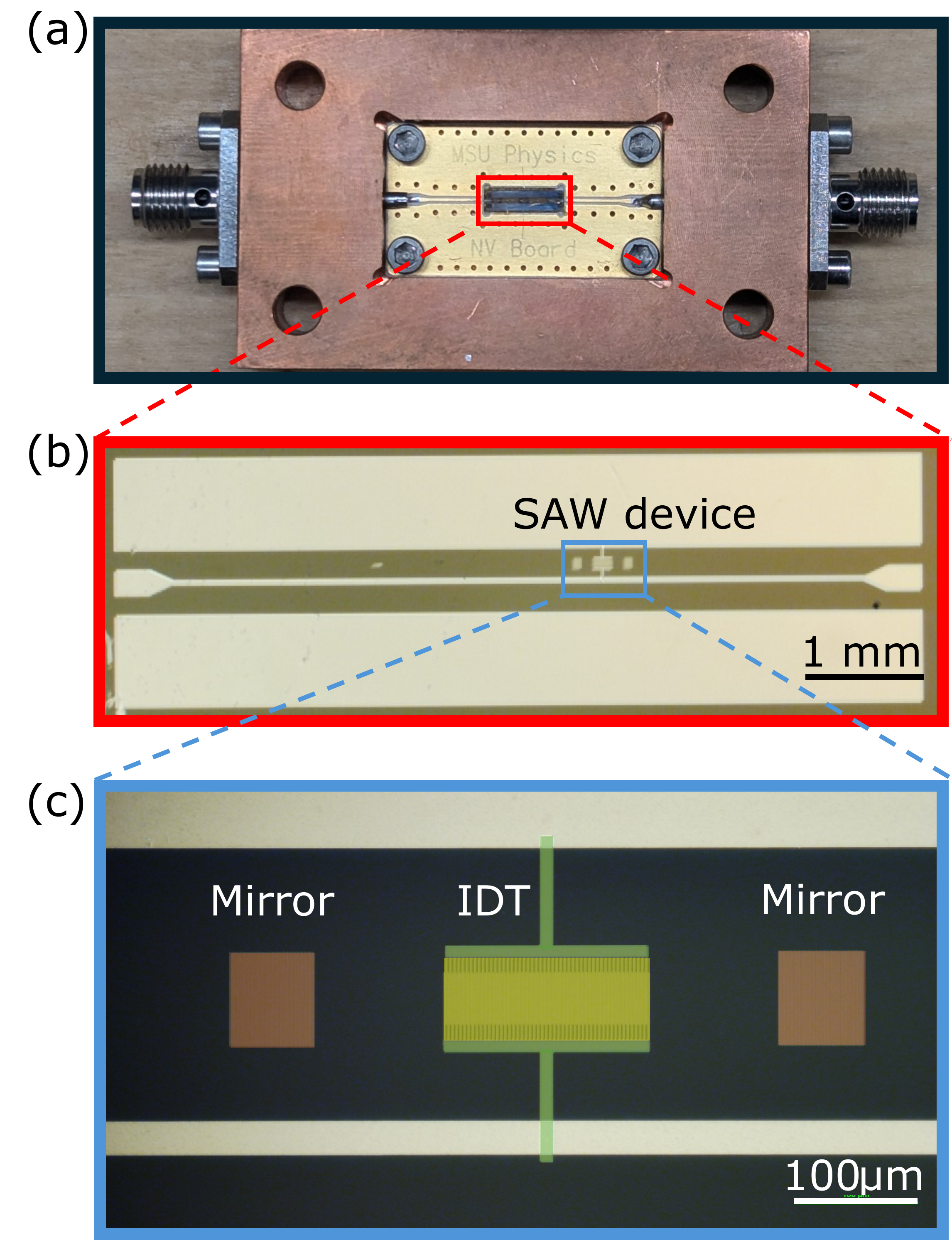}
    \caption{(a) SAW device enclosed in a copper housing with two SMA connectors. The device package includes a custom PCB featuring a central transmission line and surrounding ground planes. The transmission line on the PCB is SMA containerized on both sides, forming the input and output ports. A central rectangular slot in the PCB securely holds the SAW resonator chip. 
    (b) Zoomed-in microscope image of the device, showing the SAW resonator galvanically connected between the center trace and one of the ground planes of a CPW. 
    (c) Zoomed-in false colored microscopic image of the SAW resonator showing the central IDT (yellow) and two acoustic Bragg mirrors (brown), each $110~\mu$m from the IDT. All structures are aluminum on Y-cut LiNbO$_3$.}
\end{figure}

The device package, shown in Fig.~2~(a), features a printed circuit board (PCB) having a central transmission line and an outer ground plane, mechanically secured to a copper housing via four 2-56 screws. The transmission line is soft-soldered to the central pins of the SMA connectors on both sides, which serve as the input and output signal channels for the SAW device. A rectangular slot in the center of the PCB accommodates the lithium niobate (LiNbO$_3$) chip containing the SAW device.
The acoustic device consists of a SAW resonator galvanically connected to the central transmission line and one of the two parallel ground planes of a coplanar waveguide (CPW), as depicted in Fig.~2~(b). The on-chip CPW and PCB transmission lines are linked by aluminum wire bonds, enabling coupling with the external electrical network for measurements.
The SAW resonator, shown in Fig.~2~(c), features a central interdigitated transducer (IDT) and two acoustic Bragg mirrors positioned symmetrically on either side of the IDT, separated from the IDT by $110~\mu$m, corresponding to free propagation distance of $32$ SAW wavelengths. All on-chip components, the IDT, mirrors, transmission line, and ground planes, are fabricated from aluminum on a Y-cut LiNbO$_3$ using standard photolithography and thermal evaporation. 
The device geometry was designed~\cite{Morgan2007, kitzman2023phononic} to generate SAWs having a wavelength of $3.44~\mu$m and a multimode spectrum centered at a frequency of $1.046$~GHz.

We measured the transmission $S_{21}$ of the SAW device at room temperature using the GHFLI lock-in amplifier, as shown in Fig.~3~(a). As expected, the amplitude spectrum shows multiple resonant modes, evident as dips, highlighted by dashed lines. The corresponding phase response, after being corrected for the cable delay, is displayed in the inset. For each resonance, the phase exhibits an increase of $\sim 5^\circ$, to which a PLL will be locked later for comparison. The response in Fig.~3~(a) is equivalent to that obtained with a conventional network analyzer. Along with the distinct dips at the SAW resonances, a pronounced asymmetry in the lineshape is observed.  This is attributed to the occurrence of Fano interference, which arises from the interaction between the discrete resonant modes and continuum background states~\cite{fano1961effects}. 
In our acoustic device, this interference can arise from both phononic and electromagnetic interactions~\cite{RathFano2025, kitzman2023quantum2,Rieger2023, pitten2025effective, ren2025asymmetry}.

\begin{figure}[ht]
    \centering\includegraphics[width=1\linewidth]{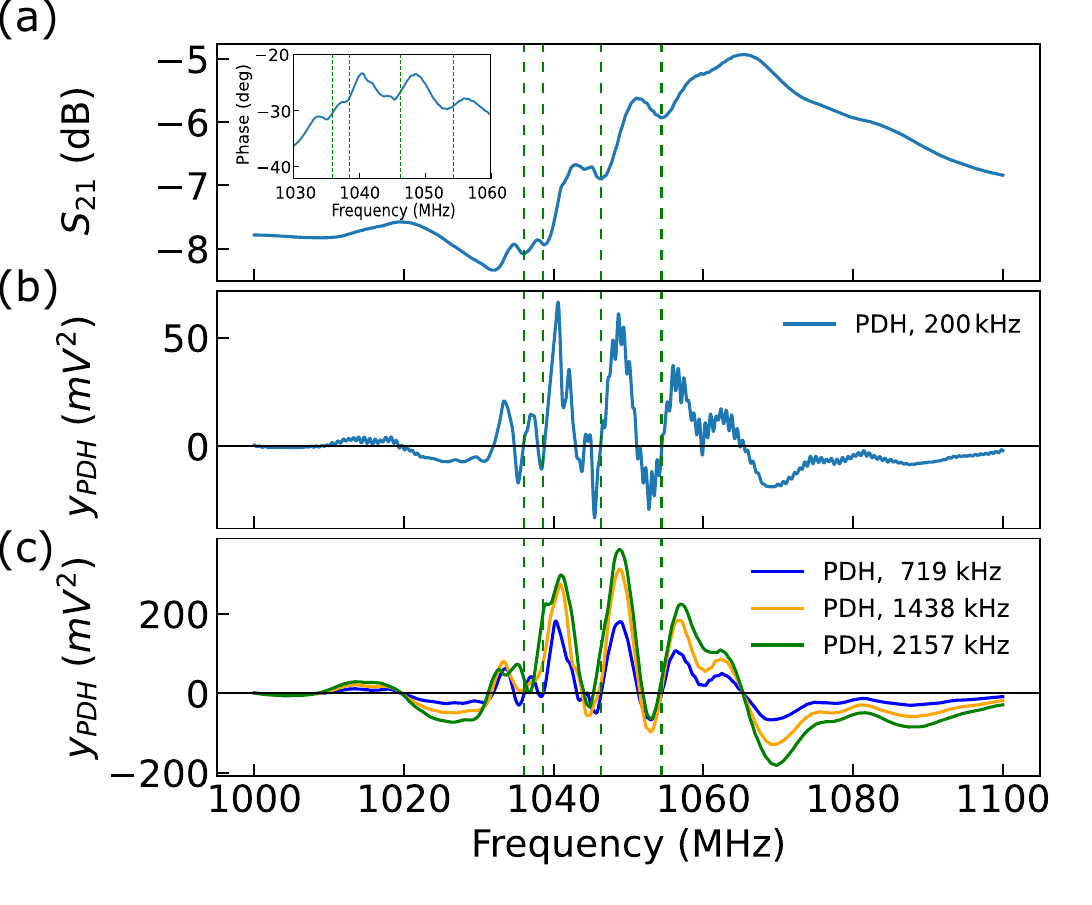}
    \caption{(a)~Transmission spectrum ($S_{12}$) of the SAW resonator using a lock-in amplifier, showing resonant dips (dashed vertical lines). Inset: Corresponding phase response. (b)~PDH observable of the SAW resonator, recorded at $f_m$ = 200 kHz, with zero crossings at the SAW resonances. The PDH measurement also reveals fine 719~kHz oscillations not apparent in the $S_{21}$ amplitude or phase response. (c)~PDH signals with ($f_m$) set to integer multiples of 719~kHz, demonstrating suppression of the spurious oscillatory features.} 
    \label{fig:spectrum}
\end{figure}

As shown in Fig.~\ref{fig:spectrum}~(b), we also measured the PDH signal from the device, where the resonator is simultaneously excited with a carrier at frequency $f$ and FM sideband tones $f\pm f_{m}$. 
The modulation frequency $f_{m}$ was $200$~kHz for this initial measurement, and $y_{PDH}$ exhibits zeros at the SAW resonances, aligning well with the dashed lines at the minima of the transmission amplitude spectrum (cf. Fig.~\ref{fig:spectrum}~(a) and (b)).
From the PDH signal (see Fig.~3~(b)), it is evident that the technique is sensitive to fine spectral features that are not obvious in either the amplitude or phase of the transmitted carrier. The most apparent feature is the periodic oscillation in $y_{PDH}$ with a period of 719~kHz. This spurious signal arises from interference between the acoustic field confined within the SAW resonator and an unintended field that leaks through the Bragg mirrors, reflects off the substrate edge, and re-enters the resonator~\cite{Morgan2007, lane2018flip}. The oscillation period is consistent with the relevant dimensions and properties of the substrate.  In sensitive experiments, such unwanted features often plague the measurement and obscure the resonator signal of interest. 
To suppress such spurious modes, acoustic dampers can be applied to the substrate, away from the resonator, to absorb the stray energy propagating outside the resonator and prevent its reflection into the device. PDH offers a convenient method for mitigating the effect of these reflected modes without incorporating any tedious mechanical modification to the device. 
By setting the modulation frequency $f_m$ to an integer multiple of 719~kHz, the separation between the carrier and sidebands matches the periodicity of the unwanted signal. Because $y_{PDH}$ is derived differentially from the three tones, oscillations with this periodicity should cancel. Figure~3~(c) shows the PDH signal recorded for $f_m = 719 \times n$~kHz, where $n=1,2,3$, confirming that the spurious features are effectively suppressed.
\begin{figure}[ht]
    \centering
    \includegraphics[width=0.95\linewidth]{4_Final_Figure4_new2.pdf}
    \caption{PLL versus PDH response to microphonic noise induced externally by mechanically tapping the connected coaxial cables. (a) PLL estimate $f_{0,PLL}$ exhibits large frequency jumps ($\approx 100$~kHz) and slow drifts, (b) PDH estimate $f_{0,PDH}$ shows smaller spikes ($\lesssim50$ kHz), rapid recovery, and negligible drift.}
\end{figure}

To directly compare the accuracy of the PLL and PDH estimates of the resonator center frequency $f_0$, we tracked the resonance frequency of the resonator using both the techniques as shown in Fig.~4. A PLL was established based on the phase of the tone at the oscillator frequency $f$, locked to the SAW resonance at $f_0~=~1054$~MHz. Therefore, the PLL estimate of the center frequency is $f_{0,PLL} = f$. Simultaneously, $y_{PDH}$ was measured using the same oscillator at $f$ along with two additional tones generated by the GHFLI's AM/FM Modulation (MOD) option at $f\pm f_m$. The three tones were demodulated at $f$ and $f\pm f_m$, and the resulting signals were used to construct $y_{PDH}$. The PDH estimate of the center frequency ($f_{0,PDH}$) was determined from $y_{PDH}$ using,
\begin{align}
f_{0,PDH}=f-y_{PDH}\frac{df}{dy_{PDH}},
\end{align}
where $\frac{df}{dy_{PDH}} = 7.3 $~kHz/m$V^2$  is the slope of $f$ \textit{vs}. $y_{PDH}$.
This slope was obtained using the PLL to switch the frequency back and forth by ~100 kHz.  We recorded the $y_{PDH}$ values and computed $\frac{df}{dy_{PDH}} = \frac{\Delta{}f} { \Delta{} y_{PDH} }.$ Here we note that the slope of $y_{PDH}$ could also be measured by sweeping the carrier frequency while keeping the modulation frequency fixed. Instead, We used the PLL as a simple and convenient way to change the carrier frequency, which allows real-time tracking of the $f_0$ while allowing us to simultaneously evaluate the $y_{PDH}$ under operating conditions.

To test the susceptibility of PLL and PDH to phase errors, we externally disturbed the system to perturb the phase of the measured signal and observed the response. A metal rod weighing $77.5$~g was dropped from a height of $0.2$~m approximately every $5$~s onto the coaxial cable connected to the resonator enclosure to induce microphonic noise~\cite{Fowler1976}. The SAW resonator itself is small and rigid, with no mechanical resonances near the induced vibration frequencies ($\leq 1$~kHz). In contrast, the coaxial cables are flexible and possess low-frequency vibrational modes. Such perturbations cause slight motion of the cables, connectors, and internal wiring within the SAW resonator enclosure, thereby modulating their parasitic capacitance. This, in turn, alters the phase of $S_{21}$ and consequently $f_{0,PLL}$. Importantly, this disturbance is expected to perturb only the phase through parasitic effects, without directly influencing the true resonant frequency $f_0$.

Fig.~4 illustrates the contrasting responses of the PLL and PDH systems to these microphonic perturbations. As expected for the phase-sensitive PLL measurement, $f_{0,PLL}$ exhibits discrete jumps of $\approx 100$~kHz, followed by slow relaxational drifts (see Fig.~4~(a)). In contrast, $f_{0,PDH}$ shows minimal drift, smaller transients ($\lesssim 50$~kHz), and significantly faster recovery from the disturbances (see Fig.~4~(b)). 
Compared to the PLL record, the PDH trace remains nearly free from frequency variations caused by the external perturbation. This behavior confirms that $f_{0,PDH}$ provides a more accurate and robust estimate of the acoustic resonator’s intrinsic center frequency $f_0$ than does $f_{0,PLL}$.

\begin{figure}[ht]
    \centering
    \includegraphics[width=1\linewidth]{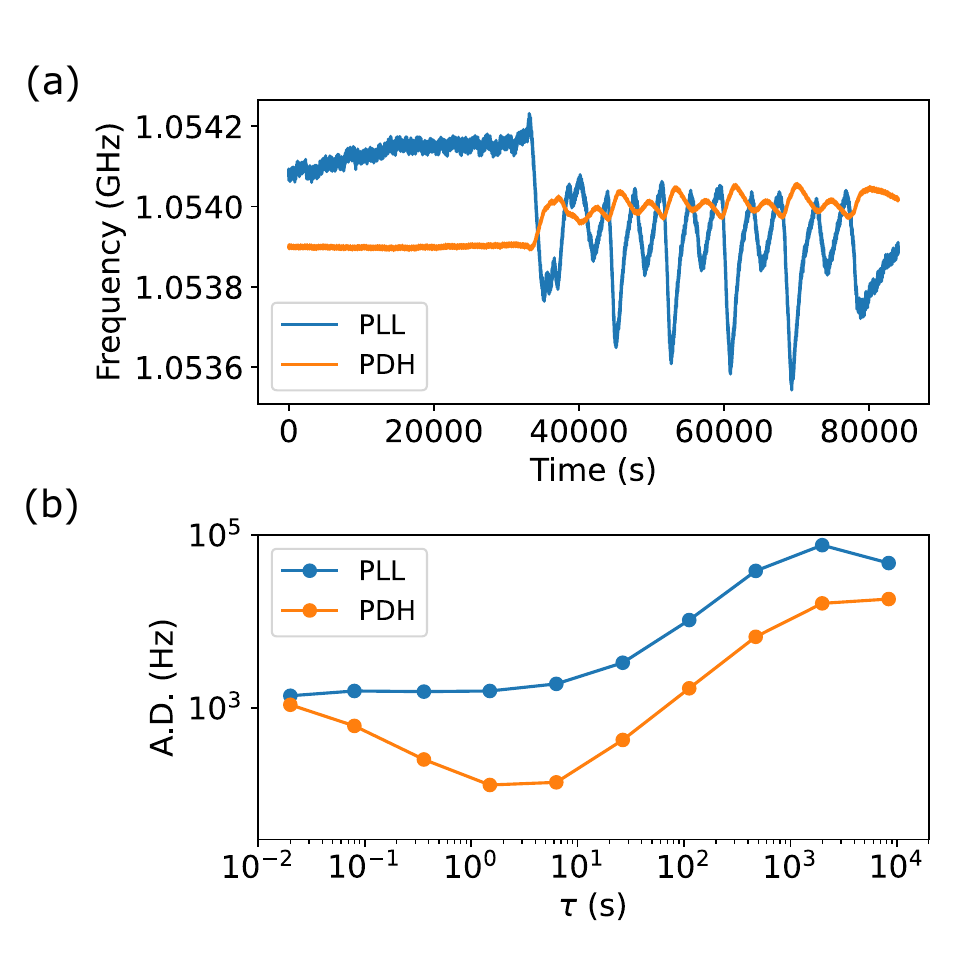}
    \caption{PLL versus PDH frequency measurement stability. (a)~Time traces of the tracked resonance frequency by PLL (blue) and PDH (orange). (b)~Allan deviation (AD) of the same data verses averaging time $\tau$, illustrating that PDH is an order of magnitude quieter than PLL for  $1\,\textrm{s}<\tau<100\,$s.} 
\end{figure}

The stability of a resonator is commonly characterized using the Allan deviation (AD)~\cite{allan1966statistics, riley2008handbook}, which quantifies the frequency stability as a function of averaging time $\tau$. The AD is usually computed over several decades in $\tau$ and is defined for a time-dependent observable $y(t)$, as,
\begin{equation}
    \sigma_y(\tau)=\sqrt{\frac{1}{2(M-1)} 
    \sum_{i=0}^{M-1} (y_{i+i}-y_i)^2
    },
\end{equation}
where $y_i=\frac{1}{\tau} \int_{i\tau}^{(i+1)\tau} y(t)\,dt$ and $M$ is the number of resonator reading intervals used to compute the $M$-sample variance.

To compare the stability of the PLL and PDH measurements, we locked a PLL to the undisturbed acoustic resonator in the same configuration described previously and simultaneously recorded the three tones required to reconstruct $y_{PDH}$. The resulting frequency records are shown in Fig.~5~(a), corresponding to a continuous 24-hour dataset, from which the AD was computed (see Fig.~5~(b)). Cyclic fluctuations on time scales exceeding $>1000$~s are evident in both the PLL and PDH time records. These oscillations are attributed to room temperature variations induced by the building’s heating, ventilation, and air conditioning (HVAC) system. In the initial period, apparently the temperature was stable without HVAC cycling.
As shown in Fig.~5(a), the PDH record exhibits substantially smaller fluctuations than the PLL record. The dependence of this difference on the averaging time $\tau$ is quantified in Fig.~5(b). The AD indicates that the PDH measurement achieves more than an order of magnitude lower noise than the PLL for $1~\textrm{s}<\tau<100~\textrm{s}$. For longer averaging times ($\tau > 100$~s), PDH remains quieter, though the difference between the two measurements becomes smaller. The lower AD observed for PDH over a wide range of $\tau$ is consistent with the fact that, unlike PLL, PDH is immune to the phase errors caused by both thermally-driven cable delay, and changes in  parasitic capacitance in cables and connectors, which perturb the phase of the resonator’s transfer function without affecting its true resonance frequency. This interpretation suggests that, for the SAW resonator, the PLL systematically overestimates frequency variations arising from both temporal and thermal fluctuations.

We note that the SAW resonator studied here is under-coupled, as seen in the lineshape presented in Figure 3 (a). While the under-coupling reduces the absolute sensitivity of the device via a reduced slope on resonance, the dominant limitation observed in the PLL configuration arises from parasitic phase drifts, which are largely suppressed by the PDH method. This observation also suggests that PDH suppresses systematic errors even in non-ideal resonator-based devices. Moreover, in the present measurement, the noise level is relatively low, and the enhanced dynamic range caused by the direct detection technique does not significantly affect the results shown in Fig. 3. However, this enhancement in dynamic range provided by the PDH method could likely become an important feature in future applications involving weaker signals or larger environmental perturbations, or for consistent measurements over a wide power range.

In summary, we have demonstrated implementation of the PDH technique on a microwave SAW resonator using a fully digital, FPGA-based lock-in amplifier. This digital realization helps in generating the frequency-modulated signal and demodulating all tones directly, which removes the dynamic range limitations and nonlinearities typically associated with analog components. 
Compared to PLL, the PDH approach provides a direct and parasitic-immune measure of the resonant frequency, suppressing errors from stray capacitance, caused by mechanical perturbation on cables. Apart from frequency stability over a long period, this technique further demonstrated a clear suppression of spurious acoustic signals by appropriate sideband tuning. These features establish PDH as a powerful alternative for high-precision resonator characterization and sensing. 
The architecture is readily extendable to audio, radio, and lower microwave frequencies and other quantum systems~\cite{adisa2025pound} . Looking forward, as SAW devices continue to be integrated into hybrid quantum systems, the PDH technique, with its immunity to parasitic noise, offers a promising pathway for quantum-limited sensing and control. 

\begin{acknowledgments}
We would like to thank C.~Undershute for valuable discussions and R.~Loloee and B.~Bi for technical assistance and use of the W.~M.~Keck Microfabrication Facility at Michigan State University. This work was supported by NSF Grant No. ECCS-2142846 (CAREER) and a Targeted Support Grant for Technology Development (TSGTD) from MSU. Additionally JP acknowledges support from the Cowen Family Endowment at MSU.
\end{acknowledgments}

\section{Author Declarations}
The authors have no conflicts to disclose.

\bibliography{Reference}

\end{document}